\documentclass [aps,prd,twocolumn,nofootinbib]{revtex4}
\usepackage{epsfig}
\usepackage{graphicx}
\usepackage{amsmath}
\usepackage{amssymb}
\usepackage{bm}
\usepackage{multirow}
\usepackage{relsize}
\begin{document}

%\preprint{}

\boldmath
\title{Global analysis of $\psi(2S)$ inclusive hadroproduction at next-to-leading order in nonrelativistic-QCD factorization}
\unboldmath
\author{Mathias Butenschoen}
\affiliation{{II.} Institut f\"ur Theoretische Physik, Universit\"at Hamburg,
Luruper Chaussee 149, 22761 Hamburg, Germany}
\author{Bernd A. Kniehl}
\affiliation{{II.} Institut f\"ur Theoretische Physik, Universit\"at Hamburg,
Luruper Chaussee 149, 22761 Hamburg, Germany}
\date{\today}
\begin{abstract}
  Working in the nonrelativistic-QCD factorization framework at next-to-leading order in $\alpha_s$, we fit the relevant color octet (CO) long-distance matrix elements (LDMEs) of the $\psi(2S)$ meson, $\langle {\cal O}^{\psi(2S)}({^1S}_0^{[8]})\rangle$, $\langle {\cal O}^{\psi(2S)}({^3S}_1^{[8]})\rangle$, and $\langle {\cal O}^{\psi(2S)}({^3P}_0^{[8]})\rangle$, to 1001 data points of upolarized and polarized $\psi(2S)$ inclusive hadroproduction.
  We do four different fits, with filters on polarization and low to middle
  transverse momentum $p_T$.
  We find that a successful description of the data is only possible with a large low-$p_T$ cut.
  Our results are one order of magnitude more precise than previous determinations of these color octet long-distance matrix elements.
\end{abstract}

\maketitle
%%%%%%%%%%%%%%%%%%%%%%%%%%%%% Text %%%%%%%%%%%%%%%%%%%%%%%%%%%%%%%%%

\section{Introduction}

In the theoretical description of heavy-quarkonium production, there is an interplay between perturbative and nonperturbative physics. A rigorous framework that aims at describing this interplay is the conjectured factorization theorem of nonrelativistic QCD (NRQCD) \cite{Bodwin:1994jh}.
Here, quarkonium production is treated as a two-step process.
In the first step, heavy quark-antiquark pairs in certain Fock states $n$, which may be color singlet (CS) or color octet (CO), are created at energy scales where perturbative calculations are feasible.
In the second step, these intermediate states then evolve into the physical heavy quarkonia, mainly via soft-gluon radiation.
This evolution is described by long-distance matrix elements (LDMEs) $\langle {\cal O}^H(n) \rangle$, nonperturbative vacuum expectation values of certain four-quark operators, specific for quarkonium $H$ and Fock state $n$.
The LDMEs obey certain rules regarding their scaling with the heavy-quark relative velocity $v$ \cite{Lepage:1992tx}.
For charmonia, $v^2\approx 0.3$ serves as a reasonably small expansion parameter.
In this work, we focus on the inclusive production of single $\psi(2S)$ mesons,
both unpolarized and polarized.
In contrast to the 1S $J/\psi$ meson and other lighter charmonia, the feed-down
from higher charmonium states is negligible here, allowing for a cleaner comparison to experimental data, which is mostly prompt, i.e.\ including such feed-down contributions.
At leading power in $v$, $\psi(2S)$ production only proceeds via the $n={^3S}_1^{[1]}$ CS state and, at next-to-leading power in $v$, the $n={^1}S_0^{[8]},{^3}S_1^{[8]},{^3}P_J^{[8]}$ CO states set in.
The corresponding CO LDMEs have to be determined from fits to data, and the goodness of these fits serves as a phenomenological test of NRQCD factorization.

Fits of the $\psi(2S)$ CO LDMEs at next-to-leading order (NLO) in $\alpha_s$ have been done previously. The fits of Refs.~\cite{Ma:2010yw,Ma:2010jj,Shao:2014yta} only included CDF data~\cite{Aaltonen:2009dm} from Tevatron run~II, with $\psi(2S)$ transverse momentum $p_T>5$~GeV (7~GeV, 11~GeV), resulting in 19 (15, 8) data points. Table 3 of Ref.~\cite{Shao:2014yta} actually features fit results for variable low-pT cut, ranging from 5~GeV to 15~GeV.
The fit of Ref.~\cite{Gong:2012ug} used the same CDF data~\cite{Aaltonen:2009dm} plus the LHCb 2012 data~\cite{Aaij:2012ag}, with $p_T>7$~GeV, amounting to 20 data points.
The more recent fit of Ref.~\cite{Bodwin:2015iua}, which incorporated resummed fragmentation function contributions in the calculation of the short-distance coefficients, included 34 data points from CDF \cite{Aaltonen:2009dm} and CMS
\cite{Chatrchyan:2011kc,Khachatryan:2015rra}.
The very recent fit of Ref.~\cite{Brambilla:2022rjd}, which assumed a relation between CO LDMEs derived using potential NRQCD~\cite{Brambilla:1999xf}, used 84 data points from Refs.~\cite{Chatrchyan:2011kc,Khachatryan:2015rra,ATLAS:2012lmu}, 25 of which refer to the $\psi(2S)$ meson.

In this work, we extend these previous analyses to many more data sets, a larger $p_T$ range, and also polarization observables, comprising a total of 1001 data points.
This allows us to reduce the errors in the LDMEs by one order of magnitude relative to the current state of the art.

This paper is organized as follows.
We outline our procedure in Sec.~\ref{sec:methods}, list our input data in
Sec.~\ref{sec:input}, present our fit results in Sec.~\ref{sec:results}, and
conclude with a summary in Sec.~\ref{sec:summary}.

\section{Method}
\label{sec:methods}

The experimental data to be fitted to come, in bins of $p_T$, as unpolarized yield $d\sigma(ab\to \psi(2S)+X)/dp_T$, with $a$ and $b$ being the colliding hadrons, and polarization observables, $\lambda_\theta$, $\lambda_\phi$, and
$\lambda_{\theta\phi}$, which appear as coefficients in the angular distribution
of the $\psi(2S)\to\mu^+\mu^-$ decay used to identify the $\psi(2S)$ meson.
Specifically, we have
\begin{align}
\frac{d\Gamma}{d\cos\theta d\phi}\propto1&+\lambda_\theta\cos^2\theta +\lambda_\phi\sin^2\theta\cos(2\phi)
\nonumber\\
& +\lambda_{\theta\phi}\sin(2\theta)\cos\phi,
\end{align}
where $\theta$ and $\phi$ are the polar and azimuthal angles of the $\mu^+$
lepton, respectively, in the $\psi(2S)$ rest frame.
The choice of coordinate axes is a matter of convention.
The three-momenta of $a$ and $b$, ${\mathbf p}_a$ and ${\mathbf p}_b$,
respectively, are generally taken to lie within the $xz$ plane, and the
various coordinate frames then differ by the choice of $z$ axis.
Specifically, the $z$ axis points along the direction of $-({\mathbf p}_a+{\mathbf p}_b)$ in the helicity (HX) frame, ${\mathbf p}_a/|{\mathbf p}_a| - {\mathbf p}_b/|{\mathbf p}_b|$ in the Collins-Soper ($\mathcal{CS}$) frame, and ${\mathbf p}_a/|{\mathbf p}_a| + {\mathbf p}_b/|{\mathbf p}_b|$ in the perpendicular helicity (PX) frame.
The observables $\lambda_\theta$, $\lambda_\phi$, and $\lambda_{\theta\phi}$ are related to the spin density matrix elements $d\sigma_{ij}$ via
\begin{gather}
\lambda_\theta=\frac{d\sigma_{11}-d\sigma_{00}}{d\sigma_{11}+d\sigma_{00}},\qquad
\lambda_\phi=\frac{d\sigma_{1,-1}}{d\sigma_{11}+d\sigma_{00}}, \nonumber\\
\lambda_{\theta\phi}=\frac{\sqrt{2}\mathrm{Re} d\sigma_{10}}{d\sigma_{11}+d\sigma_{00}}.
\label{eq:lambda}
\end{gather}
The spin density matrix elements $d\sigma_{ij}$ emerge from the unpolarized production cross sections $d\sigma$ by undoing the polarization sum and taking the polarization vectors in the amplitude and its complex conjugated counterpart to
be $\epsilon_i$ and $\epsilon_j$, respectively, in the respective reference frame.
Therefore, $\lambda_\theta$, $\lambda_\phi$, and $\lambda_{\theta\phi}$ encode the polarization information of the production process.

By NRQCD factorization, the theoretical predictions for the unpolarized cross sections and spin density matrix elements are within NRQCD factorization given by
\begin{align}
 &d\sigma_{(ij)}(ab\to \psi(2S)+X) \nonumber \\
 &= \sum_n d\tilde\sigma_{(ij)}(ab\to c\overline{c}[n]+X) \langle {\cal O}^{\psi(2S)}(n) \rangle,
\end{align}
where $d\tilde\sigma_{(ij)}(ab\to c\overline{c}[n]+X)$ are the perturbative short-distance cross sections for the production of a charm-anticharm system in Fock state $n$.
We include $n={^3}S_1^{[1]}, {^1}S_0^{[8]}, {^3}S_1^{[8]}, {^3}P_J^{[8]}$, for which
$\langle {\cal O}^{\psi(2S)}(n) \rangle$ are leading in $v$, as mentioned above.
We calculate $d\tilde\sigma_{(ij)}(ab\to c\overline{c}[n]+X)$ to NLO in $\alpha_s$ using the techniques described in Refs.~\cite{Butenschoen:2010rq,Butenschoen:2012px,Butenschoen:2019lef,Butenschoen:2020mzi}.
For the CS LDME, we use the standard choice $\langle {\cal O}^{\psi(2S)}({^3}S_1^{[1]}) \rangle=0.76$~GeV$^{-3}$, which was derived in Ref.~\cite{Eichten:1995ch} using the Buchm\"uller-Tye potential model~\cite{Buchmuller:1980su}.
Noticing that $\langle {\cal O}^{\psi(2S)}({^3}P_J^{[8]}) \rangle=(2J+1)\langle {\cal O}^{\psi(2S)}({^3}P_0^{[8]}) \rangle$, this leaves three dimensionless fit parameters,
\begin{align}
O_1 &= \langle {\cal O}^{\psi(2S)}({^1S}_0^{[8]})\rangle/\mathrm{GeV}^3,
  \nonumber\\
O_2 &= \langle {\cal O}^{\psi(2S)}({^3S}_1^{[8]})\rangle/\mathrm{GeV}^3,
  \nonumber\\
O_3 &= \langle {\cal O}^{\psi(2S)}({^3P}_0^{[8]})\rangle/\mathrm{GeV}^5,
\label{eq:oi}
\end{align}
which we determine by minimizing
\begin{align}
 \chi^2=&\sum_i \left(\frac{(d\sigma/dp_T)_i^\mathrm{data}-(d\sigma/dp_T)_i}{\Delta(d\sigma/dp_T)_i^\mathrm{data}}\right)^2 \nonumber \\
 &+ \sum_i \left(\frac{\lambda_{\theta,i}^\mathrm{data}-\lambda_\theta}{\Delta \lambda_{\theta,i}^\mathrm{data}}\right)^2 + \sum_i \left(\frac{\lambda_{\phi,i}^\mathrm{data}-\lambda_\phi}{\Delta \lambda_{\phi,i}^\mathrm{data}}\right)^2 \nonumber \\
 & + \sum_i \left(\frac{\lambda_{\theta\phi,i}^\mathrm{data}-\lambda_{\theta\phi}}{\Delta \lambda_{\theta\phi,i}^\mathrm{data}}\right)^2,
\label{eq:chi}
\end{align}
where $i$ runs over all experimental data points considered for the respective
observable.
As long as we include only data for $d\sigma/dp_T$, the fit has an analytic solution, since the theoretical predictions in the numerators depend linearly on $O_1$, $O_2$, and $O_3$.
As soon as we also include data for $\lambda_\theta$, $\lambda_\phi$, and $\lambda_{\theta\phi}$, we have to resort to numerical methods.

Assuming the Gauss distribution relation,
\begin{equation}
P(O_1,O_2,O_3) \propto e^{-\frac{1}{2}\chi^2(O_1,O_2,O_3)},
\end{equation}
between $\chi^2$ and the probability density $P(O_1,O_2,O_3)$ of the $O_i$ parameters, the inverse of the covariance matrix, defined as
\begin{equation}
 C_{ij}=\langle (O_i-\langle O_i \rangle) (O_j-\langle O_j \rangle) \rangle,
 \label{eq:covmatdef}
\end{equation}
is given by
\begin{equation}
 (C^{-1})_{ij} = \frac{1}{2}\,\frac{\partial^2 \chi^2(O_1,O_2,O_3)}{\partial O_i \partial O_j}.
\end{equation}
Equation (\ref{eq:covmatdef}) implies that the fit errors in $O_i$ are given by $\Delta O_i = \sqrt{C_{ii}}$, and the values
\begin{equation}
  V_i=\mathbf{v_i}\cdot(O_1,O_2,O_3),
\label{eq:vi}
\end{equation}
with $\mathbf{v_1}$, $\mathbf{v_2}$, and $\mathbf{v_3}$ being the three normalized eigenvectors of the real, symmetric matrix $C$, form three uncorrelated linear combinations of $O_1$, $O_2$, and $O_3$, whose uncertainties are given by the square roots of the corresponding eigenvalues.

\section{Input}
\label{sec:input}

\begin{table*}
\begin{center}
\begin{tabular}{l|ccccccccc}
 & Collab. & Year & Ref. & Collision & $\sqrt{s}$ & (Pseudo-)rapidity & $p_T$~[GeV] & Pol.\ parameters & Pol.\ frames\\
\hline
Set 1 & CDF & 2009 & \cite{Aaltonen:2009dm} & $p\overline{p}$ & 1.96~TeV & $|y|<0.6$ & 25 bins (2--30) \\
Set 2 & CDF & 1997 & \cite{Abe:1997jz}  & $p\overline{p}$& 1.8~TeV & $|\eta|<0.6$ & 5 bins (5--20) \\
Set 3 & CDF & 1992 & \cite{Abe:1992ww}  & $p\overline{p}$& 1.8~TeV & $|\eta|<0.5$ & 4 bins (6--14) \\
Set 4 & CMS & 2012 & \cite{Chatrchyan:2011kc}  & $pp$& 7~TeV & 3 bins ($|y|<2.4$) & 7--9 bins (5.5--30) \\
Set 5 & CMS & 2015 & \cite{Khachatryan:2015rra}  & $pp$& 7~TeV & 4 bins ($|y|<1.2$) & 18 bins (10--75) \\
Set 6 & CMS & 2019 & \cite{Sirunyan:2018pse}  & $pp$& 5.02~TeV & 4 bins ($|y|<0.9$) & 2--3 bins (4--30) \\
Set 7 & LHCb & 2012 & \cite{Aaij:2012ag} & $pp$ & 7~TeV & $2<y<4.5$ &  11 bins (1--16) & \multicolumn{2}{l}{(includes $\psi(2S)\to J/\psi \pi^+ \pi^-$)} \\
Set 8 & ATLAS & 2014 & \cite{Aad:2014fpa} & $pp$ & 7~TeV & 3 bins ($|y|<2$) &  10 bins (10--100) & \multicolumn{2}{l}{(uses $\psi(2S)\to J/\psi \pi^+ \pi^-$)} \\
Set 9a & ATLAS & 2016 & \cite{Aad:2015duc} & $pp$ & 7~TeV & 8 bins ($|y|<2$) &  21 bins (8--60) \\
Set 9b & ATLAS & 2016 & \cite{Aad:2015duc} & $pp$ & 8~TeV & 8 bins ($|y|<2$) &  20--24 bins (8--110) \\
Set 10 & ATLAS & 2017 & \cite{Aaboud:2016vzw} & $pp$ & 8~TeV & $|y|<0.75$ & 5 bins (10--70) & \multicolumn{2}{l}{(uses $\psi(2S)\to J/\psi \pi^+ \pi^-$)} \\
Set 11 & ALICE & 2017 & \cite{Acharya:2017hjh} & $pp$ & 13~TeV & $2.5<y<4$ & 11 bins (1--16) \\
Set 12 & ALICE & 2014& \cite{Acharya:2017hjh}  & $pp$ & 7~TeV & $2.5<y<4$ & 8 bins (1--12) \\
Set 13 & ALICE & 2016 & \cite{Adam:2015rta} & $pp$ & 8~TeV & $2.5<y<4$ & 8 bins (1--12) \\
Set 14 & CMS & 2018 & \cite{Sirunyan:2017qdw} & $pp$ & 13~TeV & 4 bins ($|y|<1.2$) & 9 bins (20--100) \\
Set 15a & LHCb & 2020 & \cite{Aaij:2019wfo} & $pp$ & 7~TeV & 5 bins ($2.0<y<4.5$) & 11 bins (3.5--14) \\
Set 15b & LHCb & 2020 & \cite{Aaij:2019wfo} & $pp$ & 13~TeV & 5 bins ($2.0<y<4.5$) & 14--17 bins (2--20) \\
Set 16 & ATLAS & 2018 & \cite{Aaboud:2017cif} & $pp$ & 5.02~TeV & 3 bins ($|y|<2$) & 5 bins (8--40) \\
Set P1 & LHCb & 2014 & \cite{Aaij:2014qea} & $pp$ & 7~TeV & 5 bins ($2<y<4.5$) & 5 bins (3.5--15) & $\lambda_\theta$, $\lambda_\phi$, $\lambda_{\theta\phi}$ & HX, $\mathcal{CS}$ \\
Set P2 & CDF & 2007 & \cite{Abulencia:2007us} & $p\overline{p}$ & 1.96~TeV & $|y|<0.6$ &  3 bins (5--30) & $\lambda_\theta$ & HX \\
Set P3 & CDF & 2000 & \cite{Affolder:2000nn} & $p\overline{p}$ & 1.8~TeV & $|y|<0.6$ & 3 bins (5.5--20) & $\lambda_\theta$ & HX \\
Set P4 & CMS & 2013 & \cite{Chatrchyan:2013cla} & $pp$ & 7~TeV & 3 bins ($|y|<1.5$) & 4 bins (14--50) & $\lambda_\theta$, $\lambda_\phi$, $\lambda_{\theta\phi}$ & HX, $\mathcal{CS}$, PX 
\end{tabular}
\end{center}
\caption{\label{tab:TheData}Overview of the data included in our fits. See text.}
\end{table*}

We evaluate the short-distance cross sections
%at NLO in $\alpha_s$ within the NRQCD factorization framework along the lines of \cite{Butenschoen:2010rq,Butenschoen:2012px,Butenschoen:2019lef,Butenschoen:2020mzi}
using the following inputs.
We take the on-shell mass of the charm quark to be $m_c=1.5$~GeV and, for definiteness, put $M_{\psi(2S)}=2m_c$ for the $\psi(2S)$ mass.
We choose the renormalization and factorization scales to be $\mu_r=\mu_f=\sqrt{p_T^2+4m_c^2}$ and the NRQCD scale to be $\mu_\Lambda=m_c$.
We adopt set CT14nlo\_NF3~\cite{Dulat:2015mca} of parton distribution functions (PDFs) from the LHAPDF~\cite{Buckley:2014ana} library, for fixed quark flavor number $n_f=3$, along with the corresponding implementation of $\alpha_s$ provided therein, which is the exact solution of the NLO renormalization group equation given by Eq.~(9.3) in Ref.~\cite{Zyla:2020zbs}, truncated after the second term. 
Furthermore, we use the branching fraction values $\mathrm{Br}(\psi(2S) \to\mu^+ \mu^-)=0.0080$, $\mathrm{Br}(\psi(2S)\to J/\psi \pi^+ \pi^-)=0.347$, and $\mathrm{Br}(J/\psi\to \mu^+ \mu^-)=0.0596$ from Ref.~\cite{Zyla:2020zbs}.

For our fits, we take into account all experimental data of $\psi(2S)$ inclusive production that we are aware of, leaving aside heavy-ion collision data, due to the large uncertainties on (or even absence of) the pertinent nuclear PDFs; total cross section data, due to the inadequacy of a fixed-order perturbative treatment at low values of $p_T$; and data on the $\psi(2S)$ to $J/\psi$ ratio of production cross sections, like the ZEUS photoproduction data~\cite{Chekanov:2002at}.
This leaves us with the proton-proton and proton-antiproton collision data listed in Table~\ref{tab:TheData}, totaling 1001 data points.
Specifically, sets 1--16
\cite{Aaltonen:2009dm,Aaij:2012ag,Chatrchyan:2011kc,Khachatryan:2015rra,Abe:1997jz,Abe:1992ww,Sirunyan:2018pse,Aad:2014fpa,Aad:2015duc,Aaboud:2016vzw,Acharya:2017hjh,Abelev:2014qha,Adam:2015rta,Sirunyan:2017qdw,Aaij:2019wfo,Aaboud:2017cif}
refer to $d\sigma/dp_T$ and sets P1--P4
\cite{Aaij:2014qea,Abulencia:2007us,Affolder:2000nn,Chatrchyan:2013cla}
to $\lambda_\theta$, $\lambda_\phi$, and/or $\lambda_{\theta\phi}$ in the HX, $\mathcal{CS}$, and/or PX frames.
The data was taken by the CDF Collaboration (sets 1--3, P2, and P3) in $p\bar{p}$ collisions at the Tevatron and by the ALICE (sets 11--13), ATLAS (sets 8--10 and 16), CMS (sets 4--6, 14, and P4), and LHCb (sets 7, 15a, 15b, and P1) Collaborations in $pp$ collisions at the LHC.
The decay channel $\psi(2S)\to \mu^+\mu^-$ is generally used for detection, except in sets 8 \cite{Aad:2014fpa} and 10 \cite{Aaboud:2016vzw}, where $\psi(2S)\to J/\psi \pi^+ \pi^-$ is used, and in set 7 \cite{Aaij:2012ag}, where both channels are used.
Note that the published data sets 7 and 11--13 each contain one further $p_T$ bin, namely $0<p_T<1$~GeV, which we exclude to avoid dealing with the infrared singularity at $p_T\to 0$, which is beyond the scope of our work.
Sets 3 and 11--13 are contaminated by non-prompt contributions, containing $\psi(2S)$ mesons from $B$ meson decays.
We correct for this by multiplying the cross section in each bin with the fraction of prompt production, which we extract from those bins in sets 2, 7, and 15b which come closest kinematically.

To investigate how the fit changes if we exclude polarized and/or low- and middle-$p_T$ production, we perform four separate fits to different subsets of the data in Table~\ref{tab:TheData}.
Fit A comprises all 1001 data points, fit B is limited to the 737 data points of unpolarized $d\sigma/dp_T$ data, fit C is limited to the 816 data points with $p_T>7$~GeV, and fit D refers to the intersection of the data samples of fits B and C, amounting to 644 points.
The specific choice of 7~GeV as the demarcation between the regions of middle
and large $p_T$ values is, of course, somewhat arbitrary and mainly for the ease of comparison with earlier fits in Ref.~\cite{Ma:2010yw,Gong:2012ug}.

\section{Results}
\label{sec:results}

\begin{table*}
\begin{tabular}{l||c|c|c|c}
 & Fit A & Fit B & Fit C & Fit D \\
\hline \hline
\multirow{2}{*}{Data fitted to} & \multirow{2}{*}{All data} & \multirow{2}{*}{All unpolarized data} & All data & All unpolarized data \\
& & & with $p_T>7$~GeV & with $p_T>7$~GeV
\\ \hline
Number of data points & 1001 & 737 & 816 & 644 \\
\hline
$O_1 = \langle {\cal O}^{\psi(2S)}({^1S}_0^{[8]})\rangle/\mathrm{GeV}^3$ & $0.000958\pm 0.000129$ & $0.0100\pm 0.0003$ & $0.00835\pm 0.00096$ & $0.0119 \pm 0.0020$ \\
$O_2 = \langle {\cal O}^{\psi(2S)}({^3S}_1^{[8]})\rangle/\mathrm{GeV}^3$ & $0.00149\pm 0.00001$ & $0.000537 \pm 0.000029$ & $0.00276 \pm 0.00012$ & $0.00225\pm 0.00025$ \\
$O_3 = \langle {\cal O}^{\psi(2S)}({^3P}_0^{[8]})\rangle/\mathrm{GeV}^5$ & $-0.000583\pm 0.000056$ & $-0.00489 \pm 0.00012$ & $0.00865 \pm 0.00055$ & $0.00612 \pm 0.00119$ \\
\hline
$\chi^2/$d.o.f. & 14.3 & 12.7 & 2.7 & 2.5 \\
\hline
Cov. matrix eigenvector $\bf v_1$ & $(0.917, -0.096, -0.387)$ & $(0.906, -0.096, -0.413)$ & $(0.867, -0.104, -0.487)$ & $(0.855, -0.107, -0.508)$ \\
Cov. matrix eigenvector $\bf v_2$ & $(0.394, 0.072, 0.916)$ & $(0.419, 0.061, 0.906)$ & $(0.497, 0.125, 0.859)$ & $(0.518, 0.121, 0.846)$ \\
Cov. matrix eigenvector $\bf v_3$ & $ (0.060, 0.993, -0.103)$ & $(0.062,0.993, -0.096)$ & $(0.029, 0.987, -0.160)$ & $(0.029, 0.987, -0.159)$ \\
\hline
$V_1={\bf v_1}\cdot (O_1,O_2,O_3)$ & $0.000962\pm 0.000141$ & $0.01103\pm 0.00030$ & $0.00275\pm 0.00110$ & $0.00680\pm 0.00234$ \\
$V_2={\bf v_2}\cdot (O_1,O_2,O_3)$ & $-0.000050\pm 0.000013$ & $-0.000200\pm 0.000014$ & $0.01192\pm 0.00013$ & $0.01161\pm 0.00014$ \\
$V_3={\bf v_3}\cdot (O_1,O_2,O_3)$ & $0.001597\pm 0.000006$ & $0.001619\pm 0.000006$ & $0.001577\pm 0.000006$ & $0.001593\pm 0.000006$ \\
Rel. errors in $\{V_1,V_2,V_3\}$ & $\{14.7\%, 26.8\%, 0.4\%\}$ & $\{2.7\%, 7.2\%, 0.4\%\}$ & $\{40.1\%, 1.1\%, 0.4\%\}$ & $\{34.4\%, 1.2\%, 0.4\%\}$
\end{tabular}
\caption{\label{tab:Fitresults}Details and results of our four $\psi(2S)$ CO LDME fits. See text.}
\end{table*}

We are now in the position to present and interpret our results.
The results of our four fits are summarized in Table~\ref{tab:Fitresults}, which, besides the obtained values of $O_i$ ($i=1,2,3$) and $\chi^2/$d.o.f., also list the eigenvectors $\mathbf{v_i}$ of the covariance matrices $C$, the linear combinations $V_i$ in Eq.~(\ref{eq:vi}), and the relative errors in the latter.
In each fit, the number of degrees of freedom (d.o.f.) is the number of data points minus three.
Notice that only the experimental errors enter the evaluation of $\chi^2$/d.o.f.\ according to Eq.~(\ref{eq:chi}).

First, we observe that all the fit results for $O_i$ in Table~\ref{tab:Fitresults} approximately obey the NRQCD velocity scaling rules~\cite{Lepage:1992tx}, a general requirement.
Second, we find that the results of fits A and B and also their qualities in terms of $\chi^2$/d.o.f.\ do not differ much, and similarly for fits C and D.
This implies that the polarization data has a limited effect on the fits.
This is not surprising, given the relatively large experimental errors in the polarization data.
Third, we find that $\chi^2$/d.o.f.\ is roughly reduced by a factor of 5 when
passing from fits A and B to fits C and D.
We thus recover the notion that a reasonably good description of the data of
$\psi(2S)$ inclusive hadroproduction by fixed-order NRQCD can only be obtained
by excluding the small-$p_T$ range with a cut of $p_T>7$~GeV or similar.

\begin{figure*}
\centering
\includegraphics[width=\linewidth]{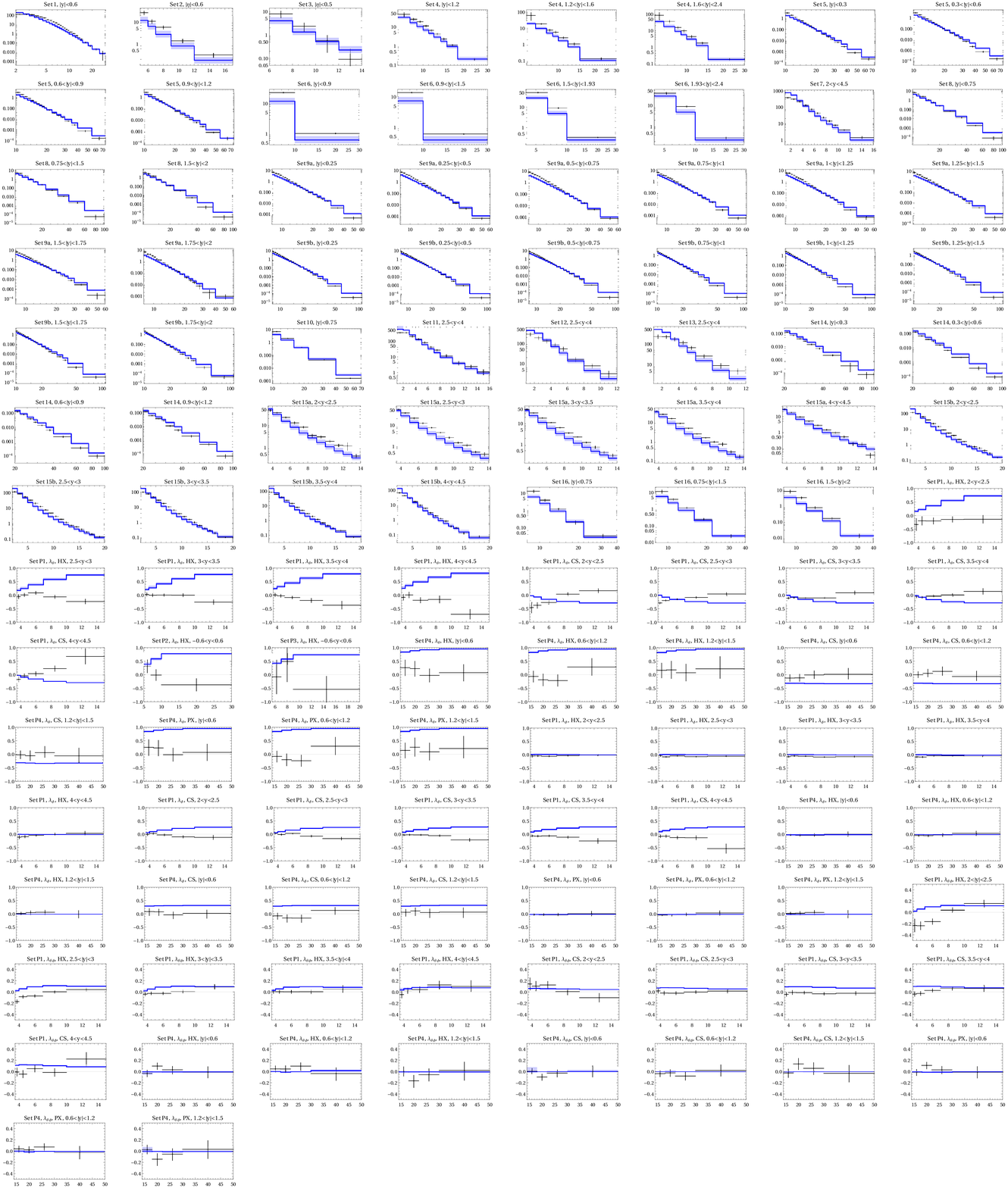}
\caption{\label{fig:FitAandData}%
The theoretical results for $d\sigma(p\overset{\text{\relsize{-1}(}-\text{\relsize{-1})}}{p}\to \psi(2S)+X)/dp_T$ [nb/GeV], $\lambda_\theta$, $\lambda_\phi$, and $\lambda_{\theta\phi}$ as functions of $p_T$ [GeV] evaluated using the results of fit A (blue) are compared to the data in Table~\ref{tab:TheData} (black).
All the shown data is fitted to.
The errors bands indicate the scale uncertainties as described in the text.}
\end{figure*}

\begin{figure*}
\centering
\includegraphics[width=\linewidth]{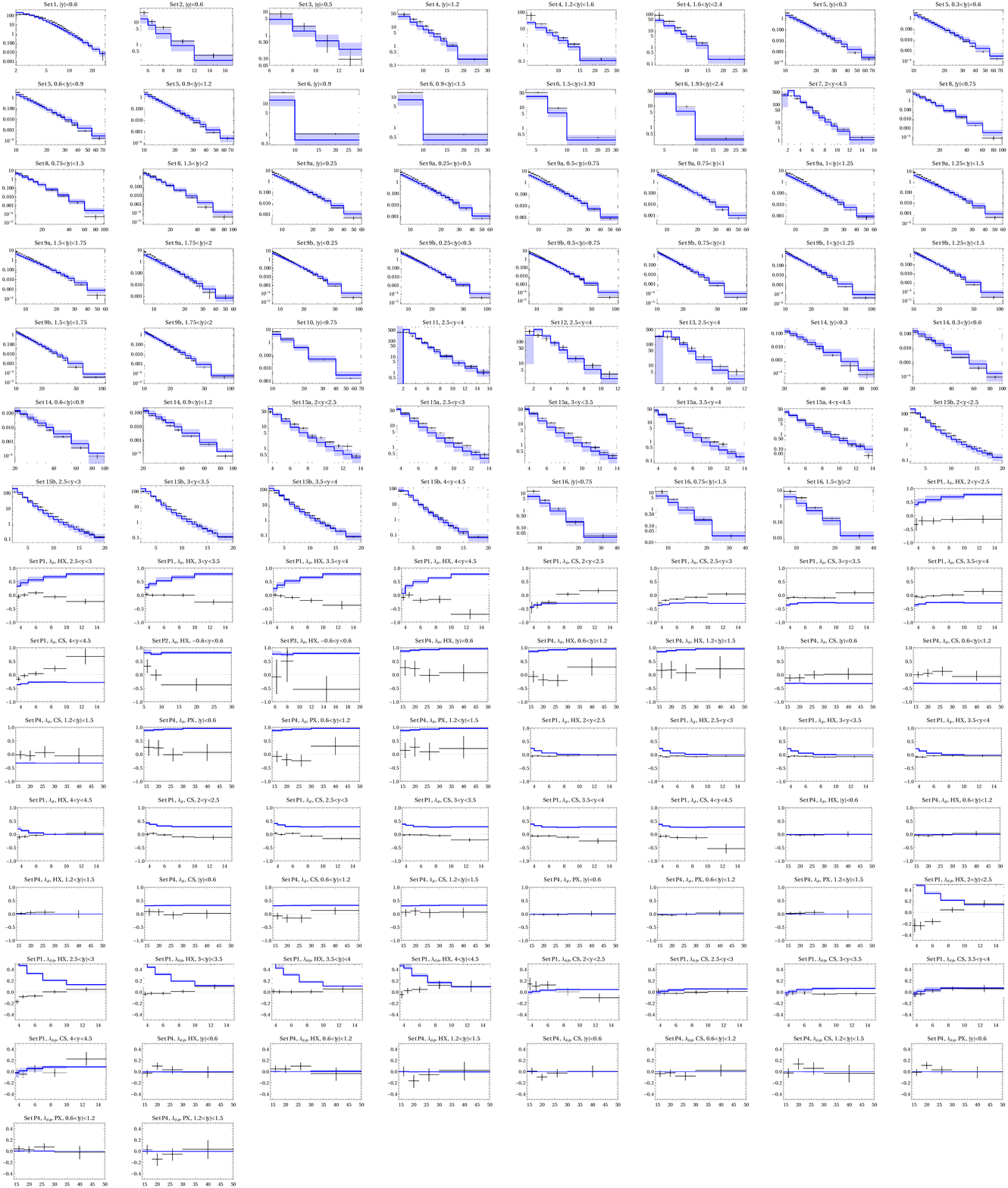}
\caption{\label{fig:FitBandData}%
As in Fig.~\ref{fig:FitAandData}, but for fit B.
Only the unpolarized data (sets 1--16) is fitted to.}
\end{figure*}

\begin{figure*}
\centering
\includegraphics[width=\linewidth]{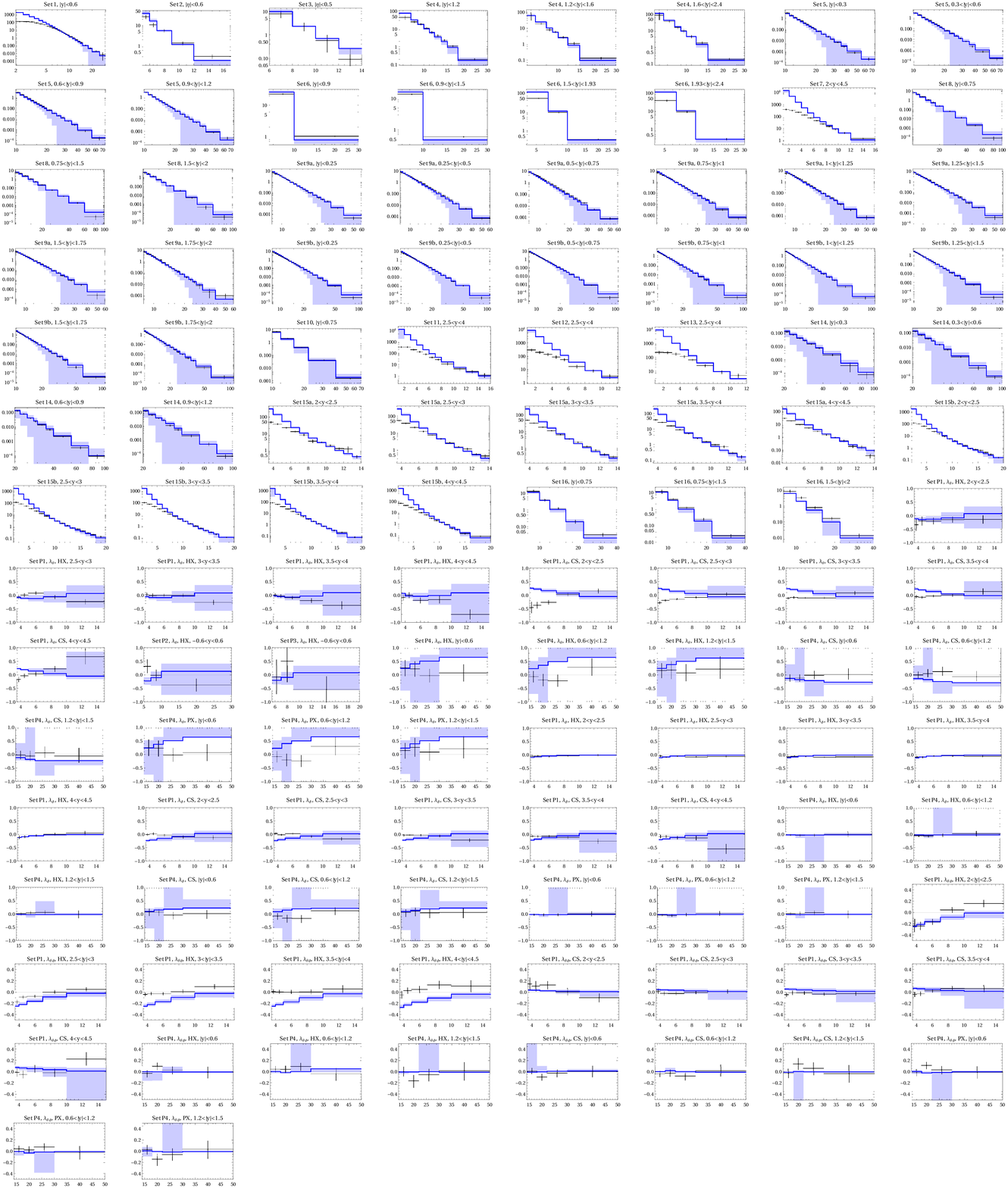}
\caption{\label{fig:FitCandData}%
As in Fig.~\ref{fig:FitAandData}, but for fit C.
Only the data with $p_T>7$~GeV is fitted to.}
\end{figure*}

\begin{figure*}
\centering
\includegraphics[width=\linewidth]{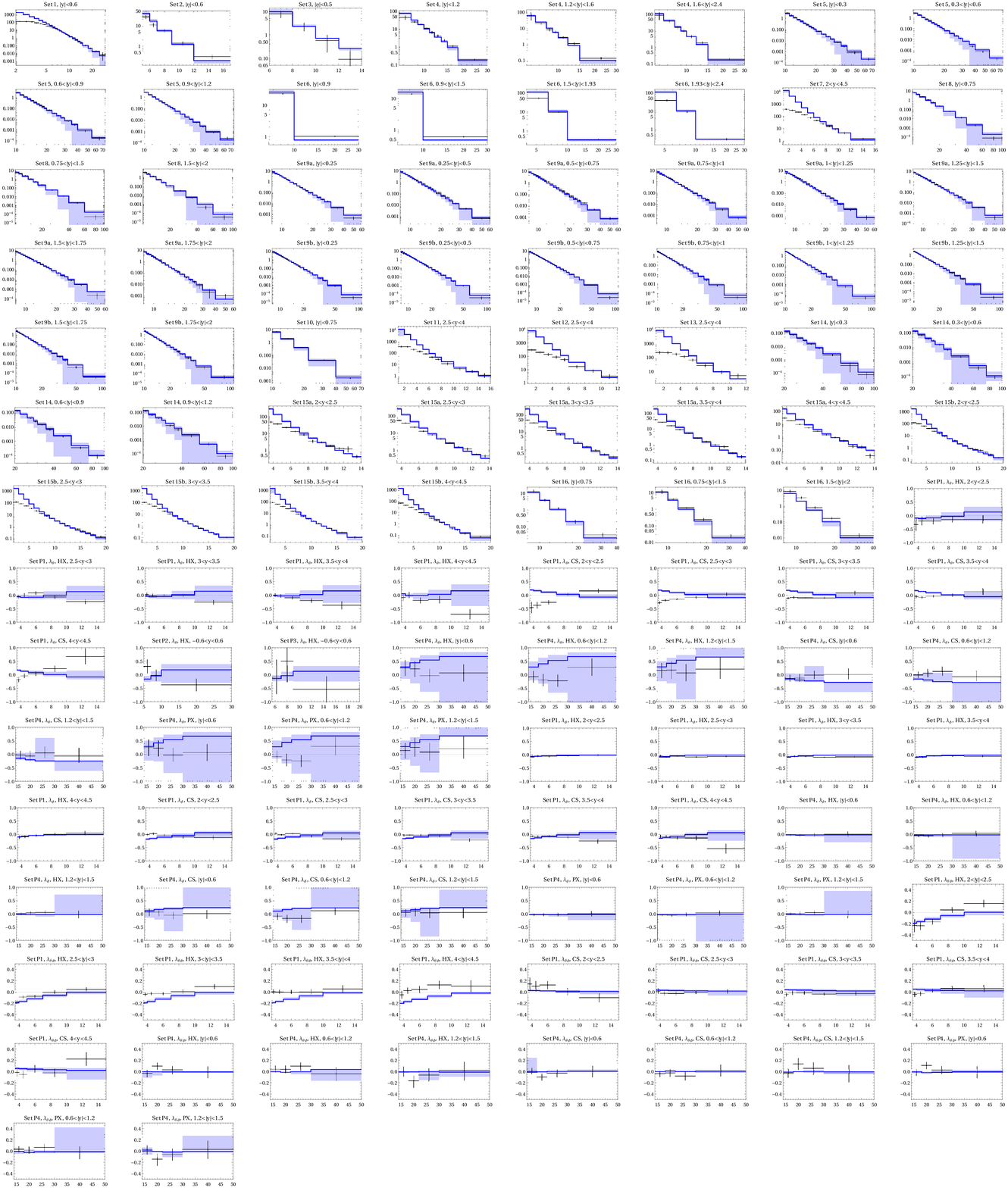}
\caption{\label{fig:FitDandData}%
As in Fig.~\ref{fig:FitAandData}, but for fit D.
Only the unpolarized data (sets 1--16) with $p_T>7$~GeV is fitted to.}
\end{figure*}

In Figs.~\ref{fig:FitAandData}--\ref{fig:FitDandData}, all the experimental data points of Table~\ref{tab:TheData} are compared with our theoretical results
evaluated using the $O_i$ values from fits A--D, respectively.
Besides the default results, also error bands are indicated, which are determined by setting $\mu_r=\mu_f=\xi\sqrt{p_T^2+4m_c^2}$ and $\mu_\Lambda=\xi m_c$ and varying the joint parameter $\xi$ between 0.5 and 2.
As in Refs.~\cite{Butenschoen:2010rq,Butenschoen:2012px,Butenschoen:2019lef,Butenschoen:2020mzi}, we implement the $\mu_\Lambda$ dependences of the LDMEs using the perturbative rather than the exact solutions of their NLO renormalization group equations, which may be found, e.g., in Eqs.~(68)--(71) of Ref.~\cite{Butenschoen:2019lef}.

Taking a closer look at Figs.~\ref{fig:FitAandData}--\ref{fig:FitDandData},
we can see where the individual fits yield good or bad descriptions of the data.
Besides slightly undershooting the unpolarized data at 4~GeV${}\alt p_T \alt 15$~GeV and slightly overshooting it at $p_T\agt 40$~GeV, fits A and B have problems describing the polarization observables.
In particular, they imply a strong transverse polarization in the HX frame, with
$\lambda_\theta\alt1$, in contrast to the largely unpolarized data,
with $\lambda_\theta\approx0$.
By contrast, fits C and D yield very good descriptions of the unpolarized data for $p_T>7$~GeV, while the data for $p_T<7$~GeV is not well described.
Fits C and D imply a significant transverse polarization in the HX frame, too,
but the tension with the data is much less pronounced than in fits A and B.
Looking at the error bands in Figs.~\ref{fig:FitAandData}--\ref{fig:FitDandData}, we observe that the results of fits A and B are very stable with respect to scale variations, while the results of fits C and D are in many regions very sensitive to scale variations, not only for the polarization observables, but also for the $d\sigma/dp_T$ distributions, where the scale choice $\xi=1/2$ even yields
negative values in the large-$p_T$ range.

\begin{figure*}
\centering
\includegraphics[width=\linewidth]{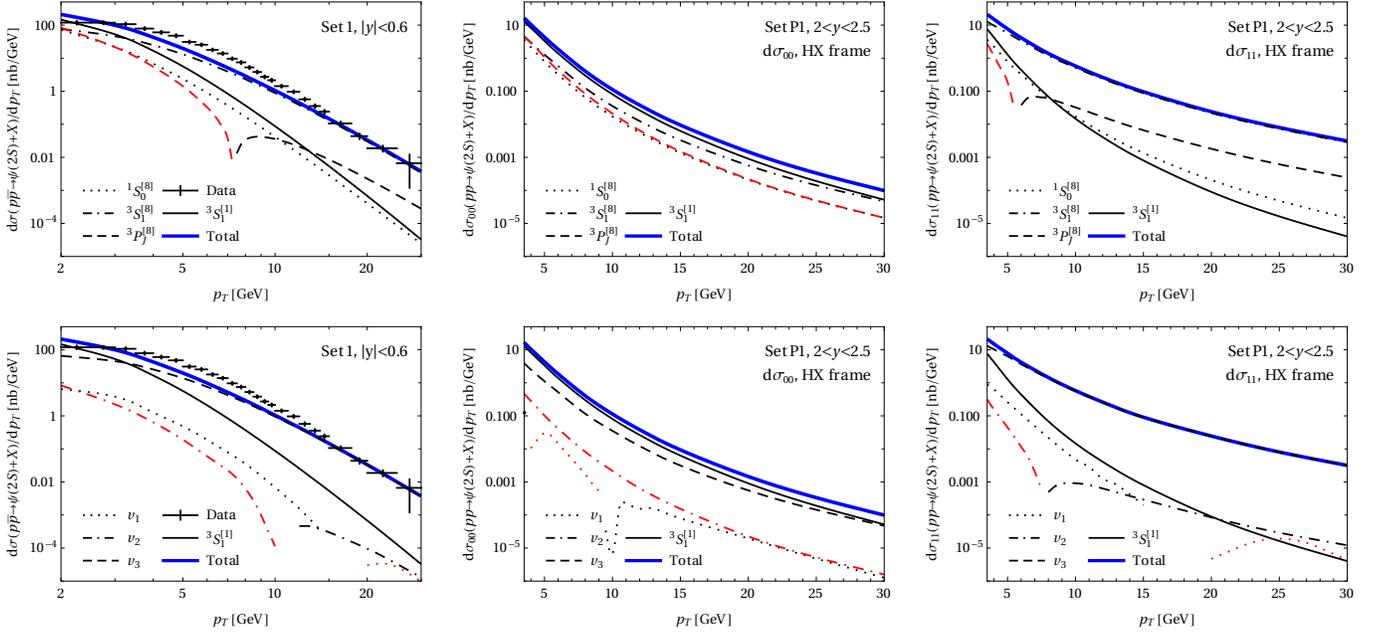}
\caption{\label{fig:DiffEigenPlotsFitA}%
  The theoretical results for $d\sigma/dp_T$ appropriate for set~1 \cite{Aaltonen:2009dm} (left column), $d\sigma_{00}/dp_T$ (center column), and $d\sigma_{11}/dp_T$ (right column) in the HX frame appropriate for the first $y$ bin of set~P1 \cite{Aaij:2014qea}, evaluated using the results of fit~A, are broken down to their CS contributions and their CO contributions proportional to $O_i$ in Eq.~(\ref{eq:oi}) (upper row) and $V_i$ in Eq.~(\ref{eq:vi}) (lower row).
Red color indicates negative values.
The data of set~1 are shown for comparison.}
\end{figure*}

\begin{figure*}
\centering
\includegraphics[width=\linewidth]{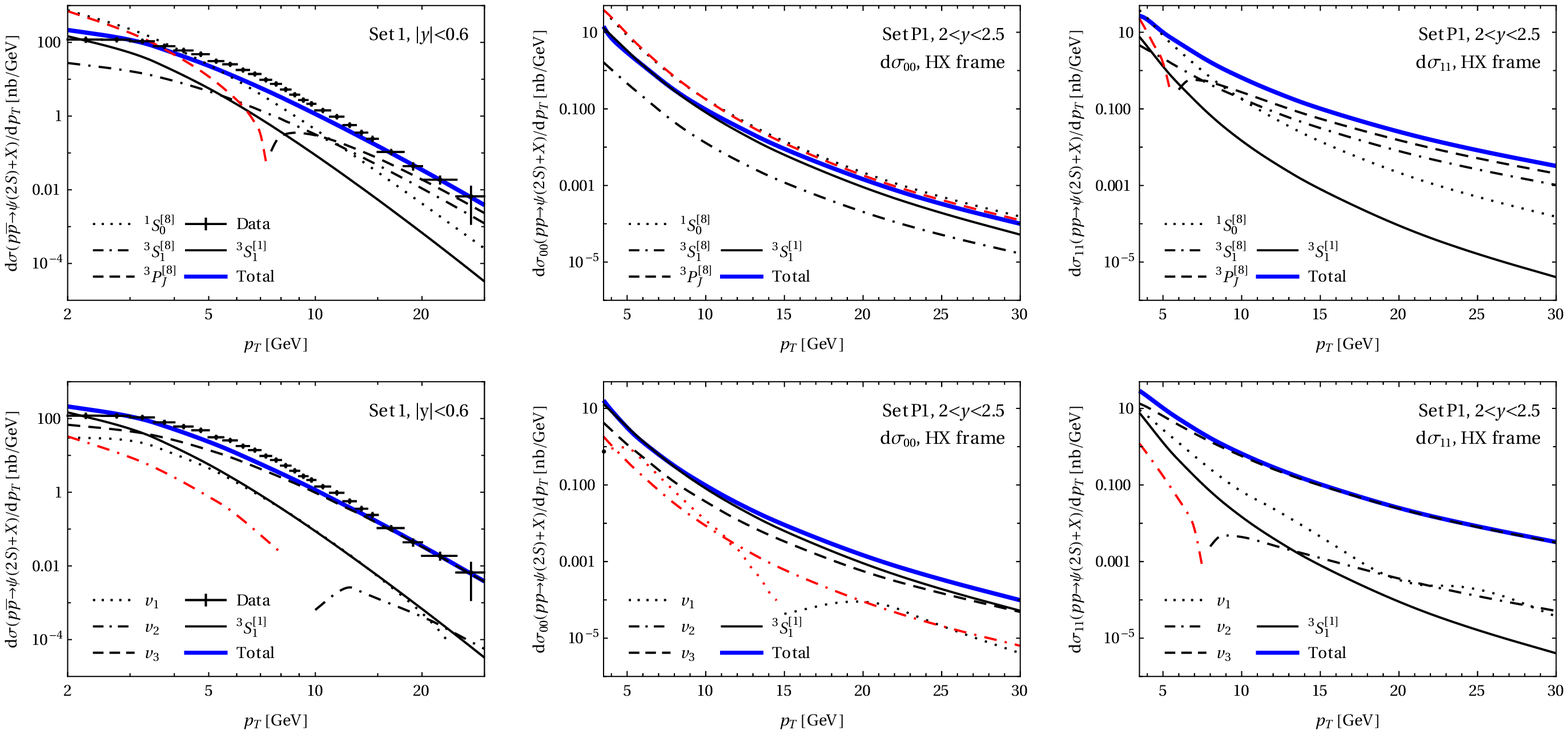}
\caption{\label{fig:DiffEigenPlotsFitB}%
Same as in Fig.~\ref{fig:DiffEigenPlotsFitA}, but for fit~B.}
\end{figure*}

\begin{figure*}
\centering
\includegraphics[width=\linewidth]{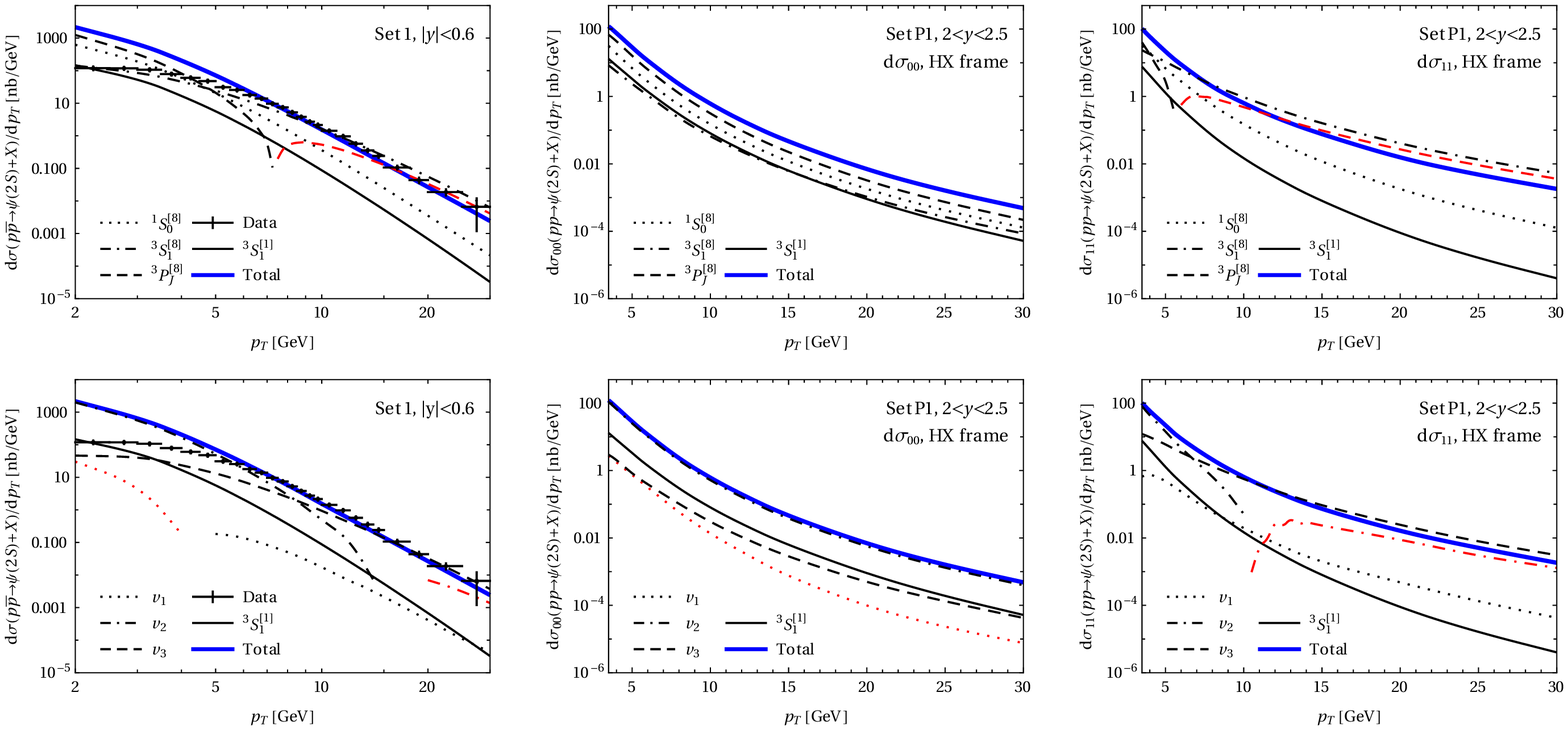}
\caption{\label{fig:DiffEigenPlotsFitC}%
Same as in Fig.~\ref{fig:DiffEigenPlotsFitA}, but for fit~C.}
\end{figure*}

\begin{figure*}
\centering
\includegraphics[width=\linewidth]{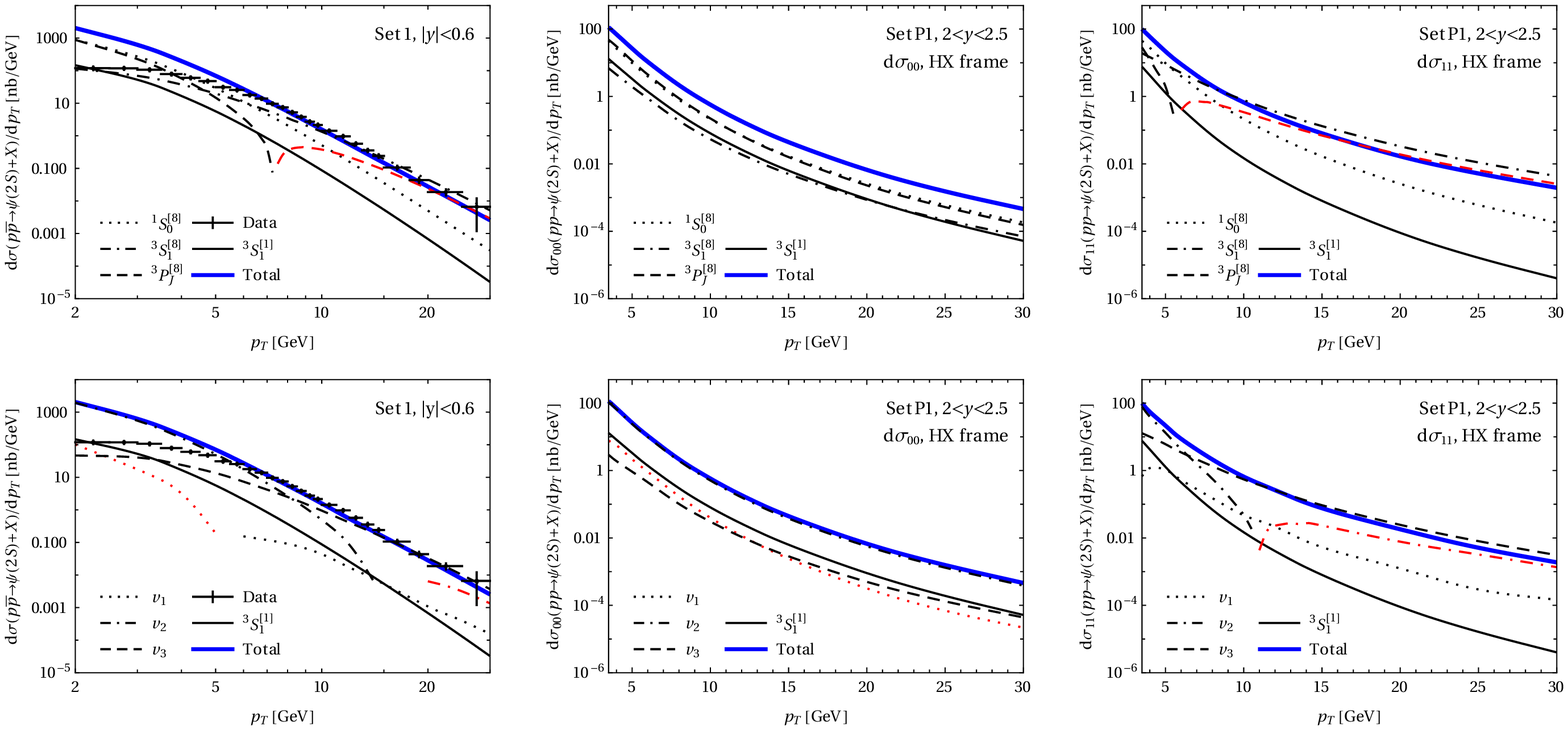}
\caption{\label{fig:DiffEigenPlotsFitD}%
Same as in Fig.~\ref{fig:DiffEigenPlotsFitA}, but for fit~D.}
\end{figure*}

In Figs.~\ref{fig:DiffEigenPlotsFitA}--\ref{fig:DiffEigenPlotsFitD}, which refer to fits A--D, respectively, we investigate the anatomy of the theoretical results for selected observables, namely, $d\sigma/dp_T$ of set~1 \cite{Aaltonen:2009dm} (first columns) and $\lambda_\theta$ in the HX frame for the lowest rapidity bin, $2<y<2.5$, of set~P1 \cite{Aaij:2019wfo}.
In fact, $\lambda_\theta$ in the HX frame is arguably the most interesting of all polarization variables because the data vs.\ theory tension has been found to be particularly prominent for it in the literature.
To achieve linearity, we actually consider $d\sigma_{00}/dp_T$ (second columns) and $d\sigma_{11}/dp_T$ (third columns) in lieu of $\lambda_\theta$, which compete for the sign of $\lambda_\theta$ according to Eq.~(\ref{eq:lambda}).
For each observable, we break down the theoretical result into the CS contribution and the CO contributions proportional to $O_i$ (upper rows) or, alternatively, to $V_i$ (lower rows).
Looking at the upper left frames of Figs.~\ref{fig:DiffEigenPlotsFitA}--\ref{fig:DiffEigenPlotsFitD}, we recover the well-known sign change of the ${^3P}_0^{[8]}$ contribution to $d\sigma/p_T$, at $p_T\approx7$~GeV for CDF kinematic conditions \cite{Butenschoen:2010rq}.
Since the short-distance cross section $\sum_{J=0}^2(2J+1)d\tilde\sigma(ab\to c\overline{c}[{^3P}_J^{[8]}]+X)/dp_T$ starts out positive at small $p_T$ values, the ${^3P}_0^{[8]}$ contributions are negative (positive) there for fits A and B (C and D) with negative (positive) $O_3$ value.
A similar feature is exhibited by $d\sigma_{11}/p_T$ in the upper right frames
of Figs.~\ref{fig:DiffEigenPlotsFitA}--\ref{fig:DiffEigenPlotsFitD}, with sign flip at $p_T\approx6$~GeV, but not for $d\sigma_{00}/p_T$ in the upper center frames.
We emphasize that individual short-distance cross sections are entitled to be negative at NLO, and this is not surprising in view of the mixing of NRQCD operators under renormalization in the $\overline{\mathrm{MS}}$ scheme \cite{Bodwin:1994jh}.
The sign flips of the $O_3$ contributions manifest themselves in appropriate
$V_i$ contributions, albeit at different $p_T$ values.
From the upper rows of Figs.~\ref{fig:DiffEigenPlotsFitA}--\ref{fig:DiffEigenPlotsFitD}, we observe that the hierarchy patterns of the $O_i$ contributions strongly depend on the fits.
The situation is quite different for the $V_i$ contributions in the lower rows of Figs.~\ref{fig:DiffEigenPlotsFitA}--\ref{fig:DiffEigenPlotsFitD}.
In fact, the $V_3$ contributions dominate for fits A and B.
As for fits C and D, the $V_2$ contributions dominate in the small-$p_T$ range
and the  $V_3$ contributions in the large-$p_T$ range.
This is also reflected by the striking smallness of the relative errors in the
respective $V_i$ values in Table~\ref{tab:Fitresults} as compared to the
residual $V_i$ values.

We may expect the results of our fit D to be compatible with those of the previous $\psi(2S)$ LDME fit in Refs.~\cite{Ma:2010yw,Ma:2010jj}, which relies on the data of set 1 \cite{Aaltonen:2009dm} with $p_T>7$~GeV.
There, the two linear combinations
\begin{align}
 M_0 =& \langle {\cal O}^{\psi(2S)}({^1S}_0^{[8]})\rangle + \frac{3.9}{m_c^2} \langle {\cal O}^{\psi(2S)}({^3P}_0^{[8]})\rangle, \\
 M_1 =& \langle {\cal O}^{\psi(2S)}({^3S}_1^{[8]})\rangle - \frac{0.56}{m_c^2} \langle {\cal O}^{\psi(2S)}({^3P}_0^{[8]})\rangle
\end{align}
were fitted.
Written in the $O_i$ basis, the corresponding vectors $\mathbf{M_0}=(0.5,0,0.87)$ and $\mathbf{M_1}=(0,0.97,-0.24)$ are indeed very close to $\mathbf{v_2}$ and $\mathbf{v_3}$ of fit D.
Also the fit results of Refs.~\cite{Ma:2010yw,Ma:2010jj}, $M_0=(0.020\pm0.006)$~GeV$^3$ and $M_1=(0.0012\pm0.0003)$~GeV$^3$, are compatible with $V_2$ and $V_3$ of fit D.
As for accuracy, $V_2$ ($V_3$) is determined by fit D 26 (67) times more precisely than $M_0$ ($M_1$) of Ref.~\cite{Ma:2010yw} and, while the third linear combination of $O_i$ could not be determined at all in Ref~\cite{Ma:2010yw}, $V_1$ is pinned down by fit D to about 34\%.

\section{Summary}
\label{sec:summary}

To summarize, working at NLO in $\alpha_s$ within the NRQCD factorization framework \cite{Bodwin:1994jh}, we have fitted the three CO LDMEs of the $\psi(2S)$ meson leading in $v$, $\langle {\cal O}^{\psi(2S)}({^1S}_0^{[8]})\rangle$, $\langle {\cal O}^{\psi(2S)}({^3S}_1^{[8]})\rangle$, and $\langle {\cal O}^{\psi(2S)}({^3P}_0^{[8]})\rangle$, to the world data of single $\psi(2S)$ inclusive hadroproduction, both unpolarized and polarized (see Table~\ref{tab:TheData}).
We have independently applied two filters to the experimental data, excluding
data with $p_T<7$~GeV and/or polarization, yielding four independent fits,
labeled A--D.
Our fit results are collected in Table~\ref{tab:Fitresults}.
They are all compatible with the velocity scaling rules of NRQCD \cite{Lepage:1992tx}.
We find that the polarization data has limited effect on the fit results, while a consistent description of all data is infeasible without a large low-$p_T$ cut, such as $p_T>7$~GeV, which reduces $\chi^2/\mathrm{d.o.f.}$ by more than a factor of 5, down to 2.7 and 2.5, leading to reasonably good overall descriptions of the data.
Thanks to the greatly enlarged data sample used, the results of our fits with $p_T>7$~GeV are one order of magnitude more precise than those of the previous fit in Refs.~\cite{Ma:2010yw,Ma:2010jj}, which is otherwise similar to ours.
This has even allowed us to pin down, with an uncertainty of about 40\%, a third linear combination of LDMEs, which has so far been out of reach.
However, the increased precision of the fits with $p_T>7$~GeV comes at the expense of reduced perturbative stability over wide kinematic ranges, manifesting itself as high sensitivity to scale variations.

At this point, one may ask how far NRQCD factorization is consolidated
or challenged in view of the advanced precision of our global analysis of single $\psi(2S)$ inclusive hadroproduction.
Unfortunately, the answer to this question is somewhat ambiguous.
While fixed-order perturbation theory, as employed here, is expected to break down in the limit $p_T\to0$ due the appearance of large soft-gluon logarithms requiring resummation, it is unclear why a small-$p_T$ cutoff as large as $p_T^{\mathrm{cutoff}}\approx2M_{\psi(2S)}$ should be necessary to enable an acceptable global fit.
In other words, one would expect smaller cutoff values, $p_T^{\mathrm{cutoff}}\lessapprox M_{\psi(2S)}$ say, to also allow for useful global fits, which they do not, as we have seen in fits A and B.
On the other hand, more serious challenges for NRQCD factorization might have just not surfaced yet, given that, in want of data, we have been confined to just one inclusive production mode, namely single hadroproduction.
This is very different for the $J/\psi$ meson, which has been observed in a variety of alternative inclusive production modes, including single photoproduction
\cite{Butenschoen:2009zy,Butenschoen:2011ks}, hadroproduction in pairs \cite{He:2015qya,He:2018hwb,He:2019qqr} or in association with a $W$ or $Z$ boson \cite{Butenschoen:2022wld}.
Furthermore, the LDMEs of the $J/\psi$ meson are related by heavy-quark spin symmetry to those of the $\eta_c$ meson, which has been observed in single inclusive hadroproduction \cite{Butenschoen:2014dra}.
It will be very interesting to also study such alternative inclusive production modes for the $\psi(2S)$ meson in the future, the more so as feed-down contributions, which complicate the $J/\psi$ case, are practically absent here.

\begin{acknowledgments}
This work was supported in part by the German Federal Ministry for Education and
Research BMBF through Grant No. 05H18GUCC1 and by the German Research Foundation DFG through Grant No.~KN~365/12-1.
\end{acknowledgments}

\end{document}